# Light Mediated Superconducting Transistor

A.V. Kavokin, I.A. Shelykh, M.M. Glazov

*LASMEA, CNRS/Université Clermont-Ferrand II Blaise-Pascal, 24, av des Landais, 63177, Aubiere, France*

**Bose-condensation of mass-less quasiparticles (photons) can be easily achieved at the room temperature in lasers. On the other hand, condensation of bosons having a non-zero mass requires usually ultra-low temperatures. Recently, it has been shown that polaritons, which are half-light-half-matter quasi-particles, may form condensed states at high temperatures (up to 300K). Polaritons composed by electron-hole pairs coupled to confined light modes in optical cavities may form a Bardeen-Cooper-Schriefer (BCS) superfluid. We propose a new transistor based on stimulated scattering of electron-hole pairs into the BCS polariton mode. A *pn*-junction embedded inside an optical cavity resonantly emits light into the cavity mode. If the cavity mode energy slightly exceeds the band-gap energy, it couples with electron-hole pairs with zero centre of mass wave-vector but non-zero wave-vector of relative motion. This creates a super-current in the plane of the structure. In an isotropic case, its direction is chosen by the system spontaneously. Otherwise, it is pinned to the external in-plane bias. We calculate the phase diagram for the electron-hole-polariton system.**

The idea to consider electron-hole pairs as Cooper pairs has been proposed a few decades ago [1]. Coulomb attraction of electrons and holes results in formation of excitons that have both fermionic and bosonic properties dependent on their binding energy and concentration. While efforts to Bose-condense excitons have yielded a few spectacular experimental works, a long-range coherence in excitonic systems appears to be very hard to create and maintain [2,3]. For the purposes of superconductivity excitons do not represent much interest as they are neutral quasi-particles. An un-bound electron-hole state can be used to create a superconducting current indeed, but it seemed that the fermionic nature of electrons and holes would prevent any pairing or condensation effects if Coulomb binding is excluded. Recently, exciton coupling to light has been shown to provide an additional source of coherence allowing for stimulated scattering and other bosonic effects in microcavities [4].

Schmitt-Rink and Chemla were among the first to point out that the BCSphase in the electron-hole liquid may be influenced by the presence of an optical field (so-called, optical Stark effect [5]). Interaction of a confined microcavity photon mode with electron-hole plasma (polariton effect) has been subject of recent theoretical research by Marchetti et al [6]. In this Letter we propose a new transistor scheme that exploits the polariton effect allowing for superconductivity in an electron-hole gas.

The scheme of proposed light-mediated superconducting transistor (LIMST) is shown in Figure 1. A *pn*-junction is embedded in an optical microcavity with thick dielectric mirrors. In order to increase the light matter coupling strength a quantum well (QW) may be embedded in the junction. From two sides it has also electrical contacts which allow to tune the density of the electron-hole gas. The cavity width is chosen in such a way that at some bias value the interband transition energy coincides with the cavity photon mode energy. In this regime, emission of photons by recombining electron-hole pairs is strongly stimulated by occupancy of the resonant cavity mode. Note that due to very thick mirrors, cavity photons may have longer life-times compared to conventional microcavities or VCSELs [1]. Remaining in the system, photons from the cavity mode can be absorbed again in the QW that creates a kind of a polariton mode. A polariton is a coherent light-matter state. It is formed if $\tau\Delta/\hbar \gg 1$, where $\tau$ is the photon mode life-time, $\Delta$ is the light-matter coupling parameter proportional to the Rabi splitting [6,7]. Note, that polaritons we discuss are created not by excitons but by un-bound electron-hole pairs. Wave-vector conservation condition allows cavity photons having a zero in-plane wave-vector to couple only with electron-hole pairs having opposite electron and hole wave-vectors $\pm k$. Such a pair coupled to light represents an elementary quantum of the super-conducting current $j_s = \frac{2}{m}ek$. Note that the binding energy of this "Cooper pair" is given by the polariton coupling constant, and the BCS gap value can be roughly estimated from the vacuum field Rabi-splitting in the system [6]. Growth of the BCS phase takes place because of stimulated scattering of the un-coupled electron-hole pairs into the polariton state. Initially, macroscopic polariton population may appear in a number of polariton states on a so-called "elastic circle", i.e. all states having the same absolute value of relative motion wave-vector and zero center-of-mass wave-vector. The number of these states is finite because of the finite lateral size of the system. At some critical density (to be estimated later), a spontaneous symmetry breaking takes place, and one of quantum states on the elastic circle becomes stronger populated than all others. At this point, a super-current appears in the plane of the structure. If an in-plane electric field is applied, the current direction will coincide with the direction of the field, while if there is no field applied any direction of current is allowed. Thus, LIMST is a device that produces a confined polariton mode in order to modulate properties of the electrical current. There is no in-coming or out-going light, which represents a fundamental difference between the LIMST and photocurrent devices and lasers.

---

[1] Vertical cavity surface emitting lasers

**Methods.**

Note, that contrary to widely discussed polariton lasers [7], LIMST operates at thermal equilibrium, and thermodynamics can be applied to estimate critical conditions of appearance of the BCS phase. Let us represent the surface density of the free energy of the system as a sum of the contributions of un-coupled electron-hole pairs having an in-plane concentration $n - n_s$ and polaritons whose concentration is $n_s$. Here $n$ is the total surface density of electron-hole pairs in the system which is conserved in the stationary regime.

The total free energy per square unit can be found as a sum of the normal and polariton contributions $F = F_n + F_s$. The polariton part is nothing but the density of energy $F_s = n_s \left( E_{ph} - \Delta \right)$, where $E_{ph}$ is an energy of the bottom of the photonic band.

The contribution of the normal part is $F_n = (n - n_s)\mu - P_n L$, where $\mu$ is a chemical potential of the electrons and holes, which is determined by the condition of conservation of the number of particles, $P_n$ is a pressure of the 2D electron-hole gas, $L$ is a thickness of the QW. If only the lowest energy subband in the QW is occupied, the expression for $F$ is easy to derive

$$F = (n - n_s)k_B T \ln\left(e^{\frac{\pi\hbar^2(n-n_s)}{mk_B T}} - 1\right) + n_s\left(E_{ph} + \varepsilon_0 - \Delta\right) + n\varepsilon_0 \quad , \tag{1}$$

where $\varepsilon_0 = \pi^2\hbar^2/mL^2$ is twice the quantum confinement energy of the electron-hole pair in the QW. Minimizing (1), one obtains an expression for the concentration of the polariton BCS phase

$$k_B T \ln\left(e^{\pi\hbar^2(n-n_s)/mk_B T} - 1\right) + \frac{\pi\hbar^2(n - n_s)}{m\left(1 - e^{-\pi\hbar^2(n-n_s)/mk_B T}\right)} = E_{ph} + \varepsilon_0 - \Delta \tag{2}$$

The solution $n_s = 0$ of Eq. (2) yields the criterion for the critical temperature $T_c$ of the BCS phase formation.

Figure 2 shows the dependence of fraction of the polariton phase on the temperature for LIMST with the concentration of carriers $n=10^{11}\,\text{cm}^{-2}$ at different values of the parameter $E_{ph} + \varepsilon_0 - \Delta$. One can see that above some critical temperature, dependent on $n$ and $E_{ph} + \varepsilon_0 - \Delta$, the polariton component completely vanishes. Between 0 and $T_c$ the dependences $n_S(T)$ have maxima. At very low temperatures, the difference between electron and hole chemical potentials is below the photon mode energy, so that the most part of electron-hole pairs remain uncoupled to light. Temperature increase helps population of the BCS phase up to some point, but then depletion of the polariton state because of thermal dissociation of the "Cooper pairs" of electrons and holes takes place. The inset of Figure 2 shows the critical temperature as a function of the total carrier concentration. For the realistic parameters of a GaAs based *pn*-junction, $T_c$ lies in the range 60-150 K. At high values of the concentration, the dependence $T_c(n)$ is almost linear and can be described by an asymptotic formula

$$k_B T_c = \frac{1.377^2}{m} n + 0.774 \left( E_{ph} + \varepsilon_0 - \Delta \right) \tag{3}$$

Reducing the detuning between the cavity photon mode and the energy gap in the *pn*-junction, one can make the critical temperature of the BCS formation lower. However, in this case, the intensity of the super-current is expected to decrease, as the contribution of each electron-hole pair to the current reduces proportionally to the decrease of the wave-vector of electron-hole relative motion. The smaller detuning, the smaller relative motion wave-vectors are selected by the cavity mode.

Here we have only taken into account coupling of the lowest energy photon state to the electron–hole plasma. This state has a zero in-plane wave-vector that is why it only couples to electrons and holes having opposite wave-vectors. Upper photonic states may couple to the pairs having non-zero centre of mass wave-vectors. Nevertheless, stimulated scattering will privilege polariton accumulation at the lowest energy state, like in polariton lasers [7].

While the absolute value of the relative motion wave-vector of the pairs in the BCS phase is fixed by the position of the cavity mode, it can have any (in-plane) direction, in general. Appearance of a directed super-current in the absence of the in-plane electric field can be considered as an evidence for the spontaneous symmetry breaking in the polariton superfluid. At thermodynamical equilibrium it may happen if the free energy of a state with the broken symmetry is lower than the free energy of a symmetrical state.

Let us consider the polariton phase with the total number of particles $N_S$, $N_0$ of which are situated at the single splitted state and $N_1$ are situated at the elastic circle having $M$ quantum states in total. In the first order of perturbation theory the energy shift due to the interactions of electron-hole pairs in the condensate is

$$\Delta E = V(0) \left[ N_0^2 + \frac{N_1^2}{M} \right], \tag{4}$$

where $V(0)$ is the matrix element of polariton-polariton interaction for the pair of polaritons occupying the same quantum state. It is dominated by dipole-dipole interaction of electron-hole pairs and has a negative sign, therefore. The entropy of the system is

$$S = k_B N_1 \ln \frac{M + N_1}{N_1}, \quad (5)$$

which yields the free energy

$$F(N_1) = V(0)\left[(N_s - N_1)^2 + \frac{N_1^2}{M}\right] - k_B T N_1 \ln \frac{M + N_1}{N_1}. \quad (6)$$

Figure 3 shows the dependencies of the free energy on $N_0/N_s$ for different temperatures. We have taken $N_s = 5 \cdot 10^4, V(0) = -2 \cdot 10^{-7} eV$. At zero temperature the free energy is a monotonously decreasing function of $N_0$ and its minimum is reached when all the polaritons are in the splitted state (curve (a) in Figure 3). At higher temperatures depletion of the splitted state takes place and the free energy minimum gradually shifts towards $N_0 = 0$. At the temperature about 131 K the minimum of the free energy vanishes, and polariton distribution in the reciprocal space becomes isotropic. The inset to Figure 3 shows the phase diagram for the model LIMST. We distinguish between the superfluid phase (1) given by a condition $N_0 > 0$, isotropic polariton phase ($N_0 = 0, n_s > 0$), and the normal phase $n_s = 0$. One can see that for high enough concentrations of carriers the critical temperature of the superfluide transition may approach room temperature.

The present model is oversimplified indeed. It neglects various mechanisms of dephasing in the system, which is justified only in case when the pair binding parameter $\Delta$ exceeds substantially $/\tau_d$, where $\tau_d$ is a characteristic dephasing time. As usually $\tau_d$ decreases with the temperature increase, it imposes an additional constraint on the critical temperature $T_c$. In order to realise a LIMST working at room temperature, one should chose materials with enhanced light-matter coupling strength. Wide-band gap semiconductors like GaN or ZnO seem to be good candidates from this point of view.

We thank F. Marchetti, M. Szymanska and G. Malpuech for stimulating discussions. This work has been supported by the Marie Curie MRTN-CT-2003-503677 "Clermont2".

# Figure captions

Figure 1.     Scheme of the LIMST device: a *pn*-junction is embedded inside a microcavity. Resonance of the interband transition with the cavity mode leads to formation of a superfluid current.

Figure 2. Dependence of the BCS fraction in the electron-hole plasma on temperature for different values of $E_{ph}+\varepsilon_0-\Delta$. Curves 0,1,2,3 correspond to $E_{ph}+\varepsilon_0-\Delta=0$, 1, 3, 5 meV, respectively. The inset shows critical temperature versus electron-hole pairs concentration. Curves 1,2,3 correspond to $E_{ph}+\varepsilon_0-\Delta=1, 3, 5$ meV, respectively.

Figure 3. Free energy of the polariton condensate as a function of the fraction of polaritons in the splitted state at (a) $T=0\,\text{K}$, (b) $T=120\,\text{K}$, (c) $T=135\,\text{K}$. The inset shows the phase diagram of the LIMST, the area 1 corresponds to the superfluid phase, the area 2 shows the isotropic polariton phase, the area 3 shows the normal phase.

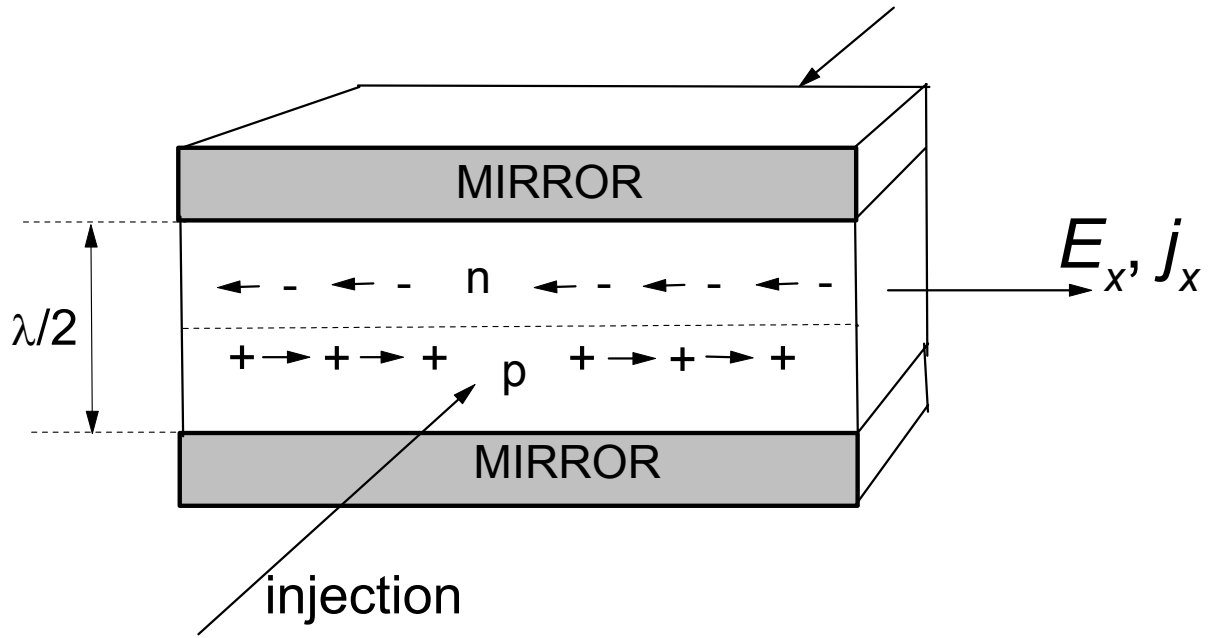

Kavokin et al, Figure 1

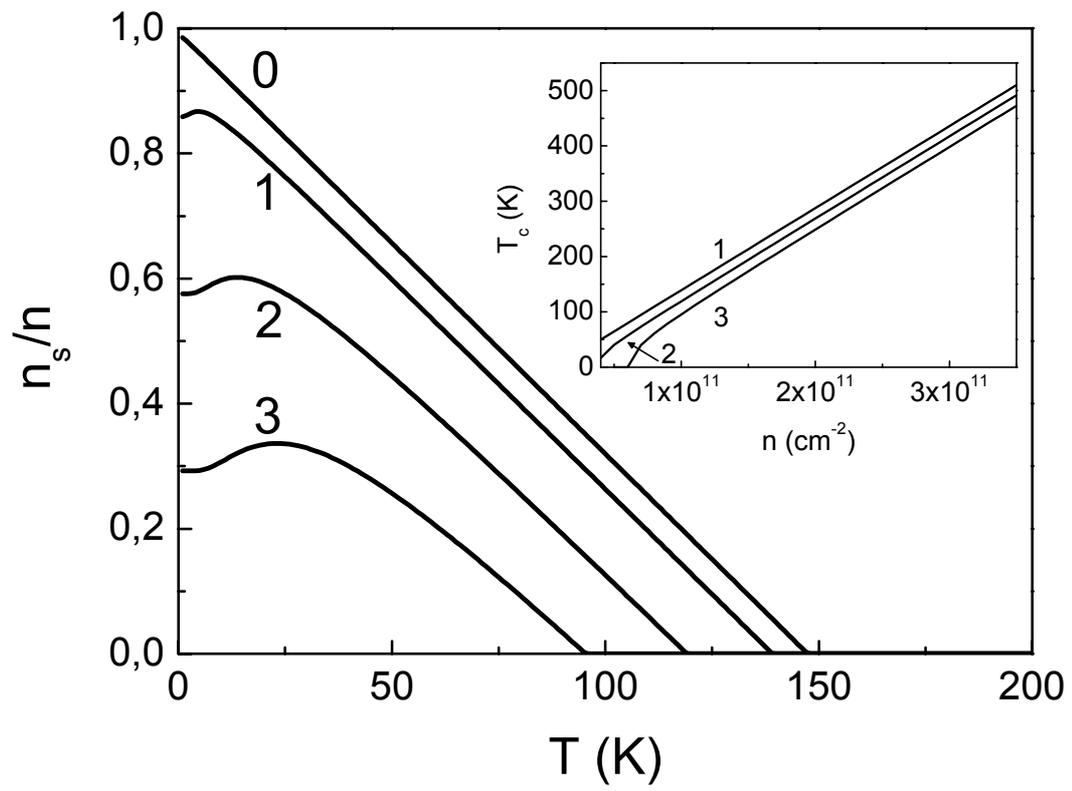

Kavokin et al, Figure 2

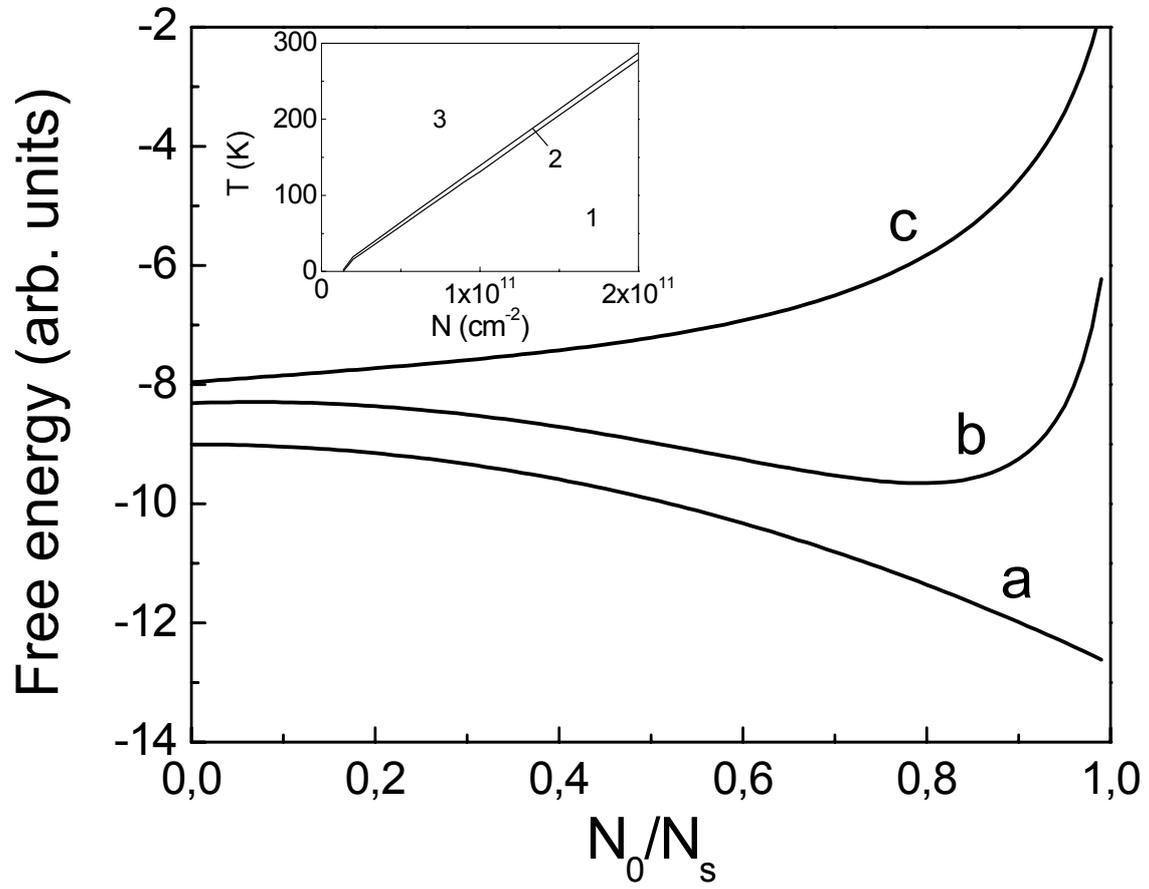

Kavokin et al. Figure 3

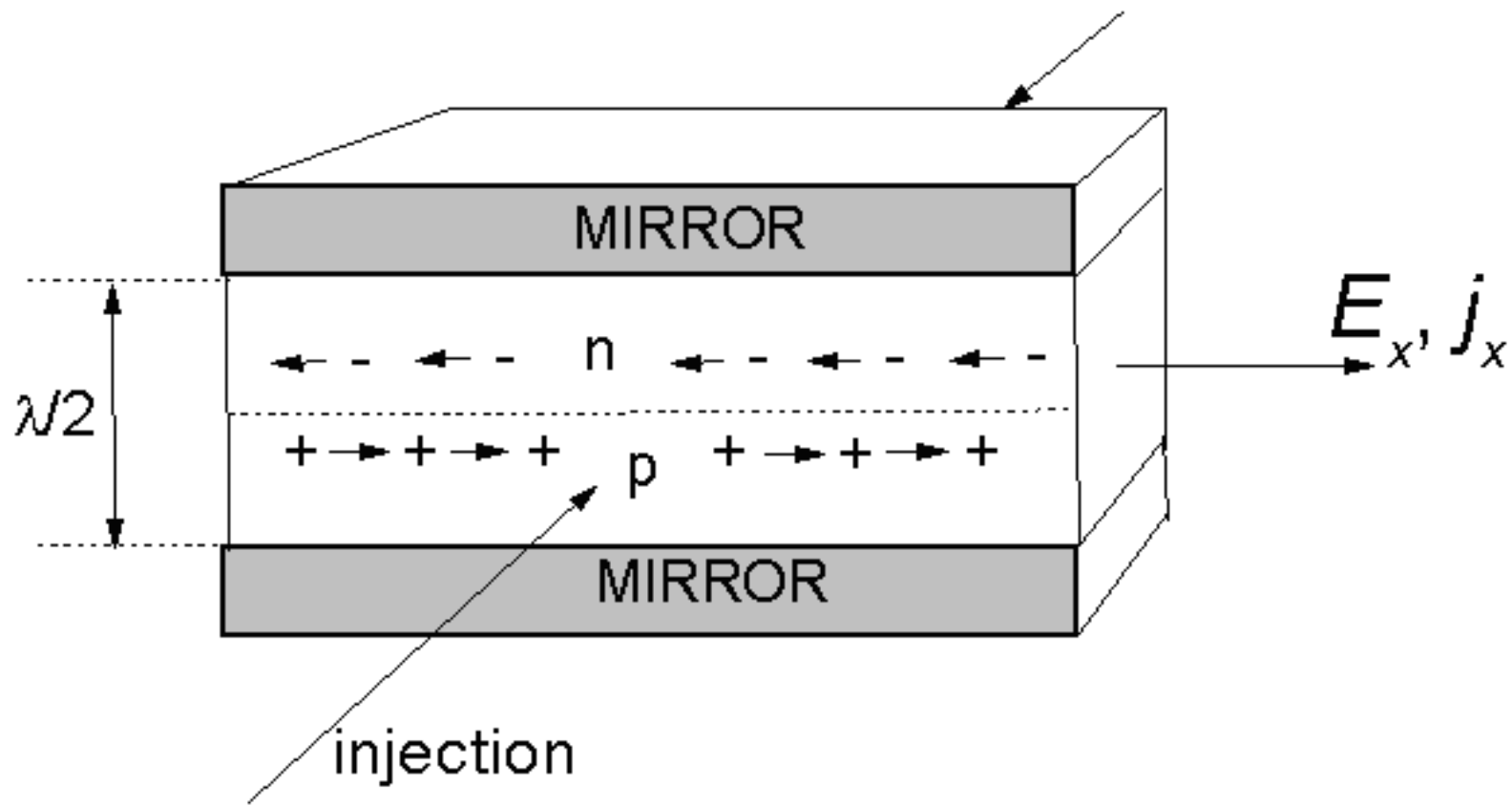

Kavokin et al, Figure 1

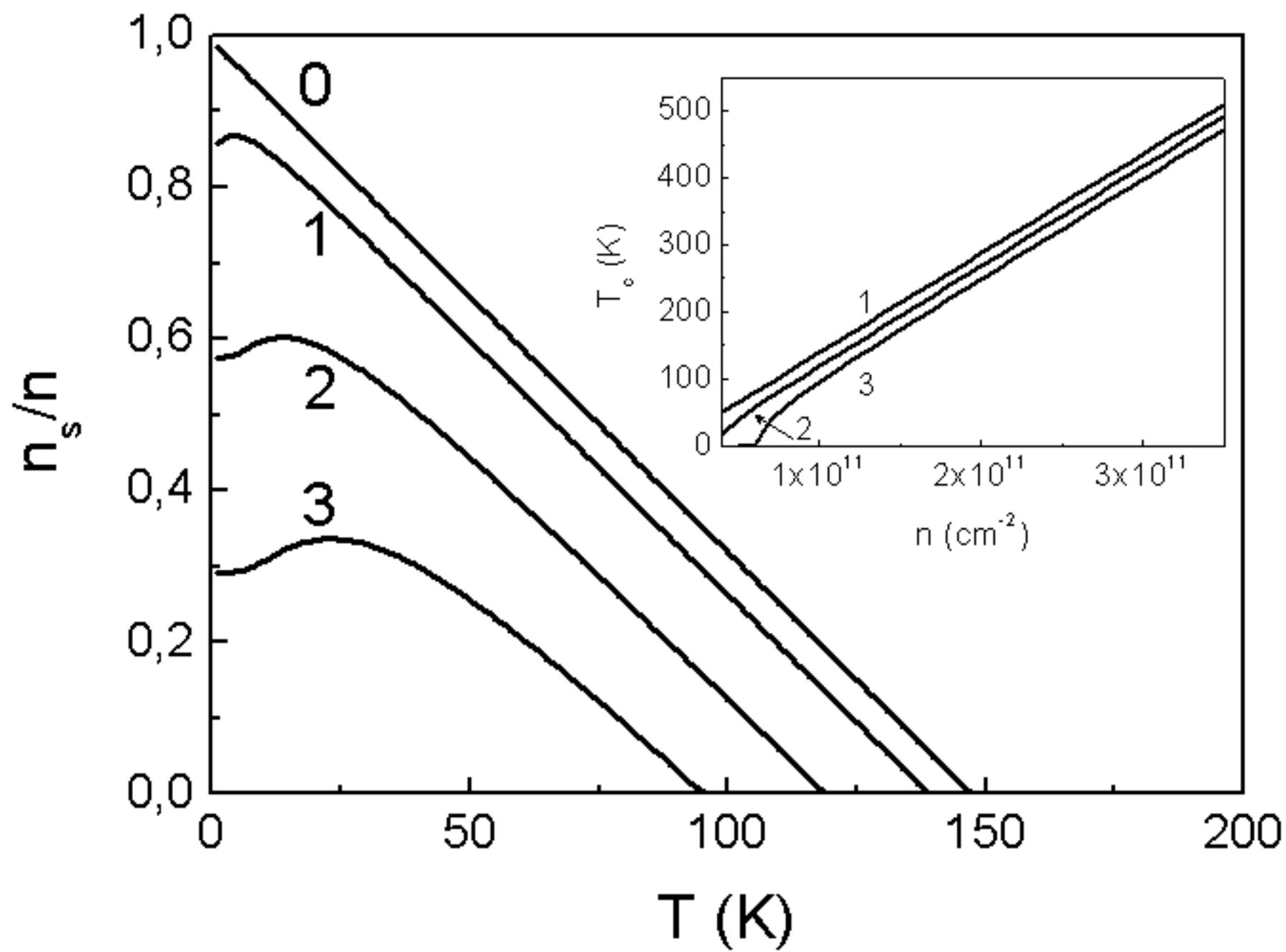

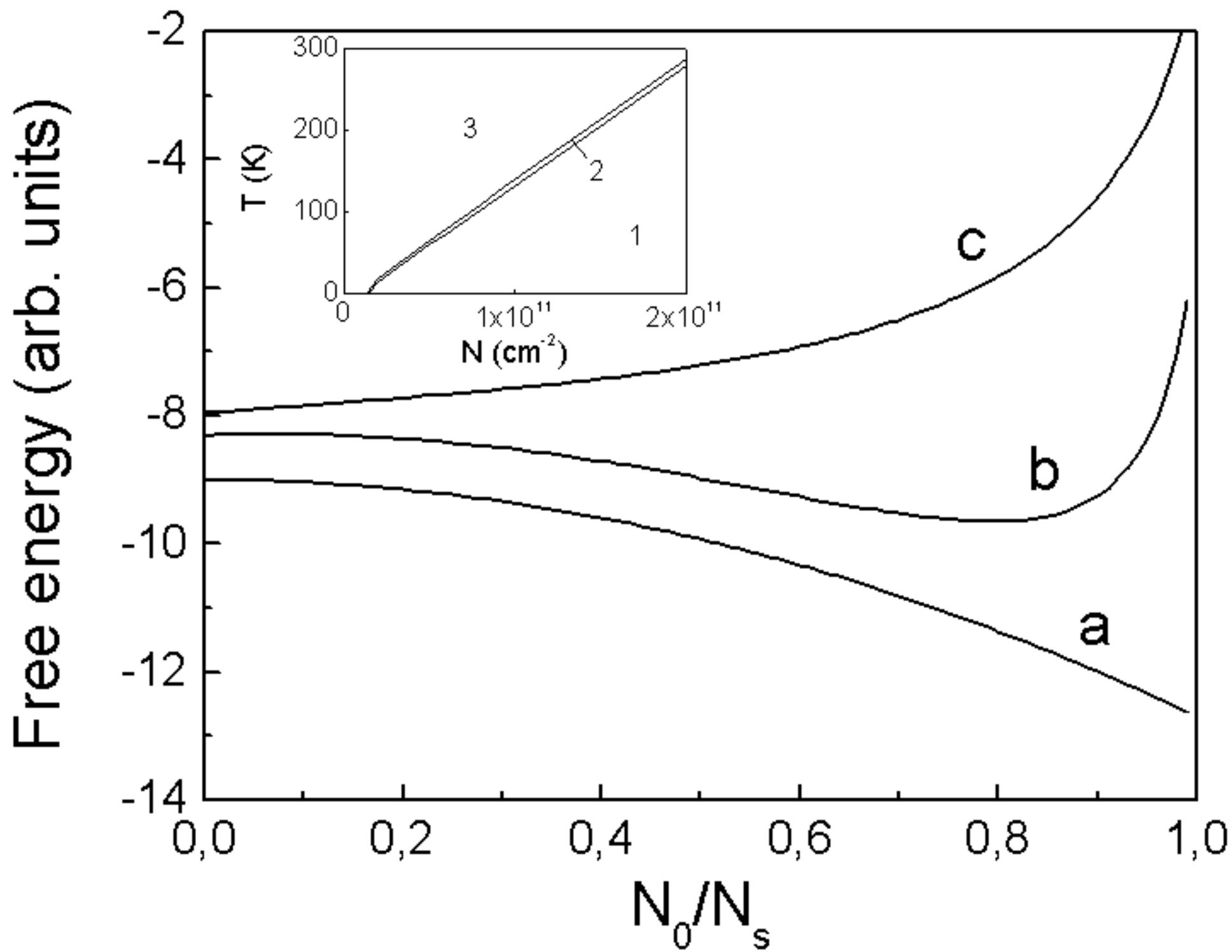